# Heterogeneous Relational Databases for a Grid-enabled Analysis Environment

Arshad Ali[1], Ashiq Anjum[1,4], Tahir Azim[1], Julian Bunn[2], Saima Iqbal[2,4], Richard McClatchey[4], Harvey Newman[2], S. Yousaf Shah[1], Tony Solomonides[4], Conrad Steenberg[2], Michael Thomas[2], Frank van Lingen[2], Ian Willers[3]

*[1]National University of Sciences & Technology, Rawalpindi, Pakistan*
*Email: {arshad.ali, ashiq.anjum, tahir, yousaf.shah}@niit.edu.pk*
*[2]California Institute of Technology, Pasadena, USA*
*Email: {Julian.Bunn, fvlingen}@caltech.edu, {newman, conrad, thomas}@hep.caltech.edu*
*[3]European Organization for Nuclear Research, Geneva, Switzerland*
*Email: {Ian.Willers, Saima.Iqbal}@cern.ch*
*[4]University of the West of England, Bristol, UK*
Email: {Richard.McClatchey, Tony.Solomonides}@uwe.ac.uk

## Abstract

*Grid based systems require a database access mechanism that can provide seamless homogeneous access to the requested data through a virtual data access system, i.e. a system which can take care of tracking the data that is stored in geographically distributed heterogeneous databases. This system should provide an integrated view of the data that is stored in the different repositories by using a virtual data access mechanism, i.e. a mechanism which can hide the heterogeneity of the backend databases from the client applications.*

*This paper focuses on accessing data stored in disparate relational databases through a web service interface, and exploits the features of a Data Warehouse and Data Marts. We present a middleware that enables applications to access data stored in geographically distributed relational databases without being aware of their physical locations and underlying schema. A web service interface is provided to enable applications to access this middleware in a language and platform independent way. A prototype implementation was created based on Clarens [4], Unity [7] and POOL [8]. This ability to access the data stored in the distributed relational databases transparently is likely to be a very powerful one for Grid users, especially the scientific community wishing to collate and analyze data distributed over the Grid.*

## 1. Introduction

In a geographically distributed environment like the Grid, database resources can be very diverse because they are developed by different vendors, run on different operating systems, support different query languages, possess different database schemas and use different technologies to store the same type of data. Furthermore, this data is accessed by applications developed for different platforms in varying development environments. Currently our Grid environment is using two major formats for the storage of data: file-based data and relational databases. Sophisticated systems are in place to track and manage files containing data and replicated at multiple storage sites. These systems generally rely on cataloging services to map file names with their physical storage locations and access mechanism. By finding out the physical location of a file and its access protocol from the catalog, the user can easily gain access to the data stored in the file. Examples of cataloging services in use for this purpose are the Replica Location Service (European DataGrid (EDG) [1] and Globus [2]), and the POOL File Catalog [3].

However, besides file-based data, significant amounts of data are also stored in relational databases. This is especially true in the case of life sciences data, astronomy data, and to a lesser extent, high-energy physics data. Therefore, a system that provides access to multiple databases can greatly facilitate scientists and users in utilizing such data resources.

Software for providing virtualized access to the data, in a similar way to files, is only beginning to appear. Most data grids currently do not incorporate much support for data stored in multiple, distributed databases. As a result, users have to send queries for data to each of the databases individually, and then manually integrate the returned data. This integration of data is essential in order to obtain a consistent result of a query submitted against the databases.

In this paper, we present a system that has been developed to provide Grid users efficient access to globally distributed, relational databases. In our prototype, the crucial issue of integration of data is addressed in three stages: initially data is integrated at the data warehouse level where data is extracted from the normalized schema of the databases, and loaded into the denormalized schema of the data warehouse. Secondly, materialized views of the data are replicated from the data warehouse and stored in data marts. Finally, data generated from the distributed queries, which run over the data marts, is integrated and these integrated results are presented to the clients. A Clarens [4] based web service interface is provided in order to enable all kinds of (simple and) complex clients to use this prototype over the web conveniently. Furthermore, in order to avoid the performance issue of centralized registration of data marts and their respective schema information, the Replica Location Service (RLS) is used.

The paper is organized as follows. In Section 2, we briefly describe the background of the issue addressed, and describe the requirements in detail. In Section 3, we give a brief overview of previous related work in this direction, before plunging into a full description of the architecture and design of the system in Section 4. Performance statistics are presented in Section 5, and the current status of the work and possible extensions for the future are mentioned in Section 6. We finally conclude our discussion in Section 7.

## 2. Background

The Large Hadron Collider (LHC) [5], being constructed at the European Organization for Nuclear Research (CERN) [6], is scheduled to go online in 2007. In order to cater for the large amounts of data to be generated by this enormous accelerator, a Grid-based architecture has been proposed which aims to distribute the generated data to storage and processing sites. The data generated is stored both in the form of files (event data) and relational databases (non-event data). Non-event data includes data such as a detector's calibration data and conditions data.

While sophisticated systems are already in place for accessing the data stored in distributed files, software for managing databases in a similar way is only beginning to be developed. In this paper, we propose a web service based middleware for locating and accessing data that is stored in data marts. A prototype was developed based on this proposed middleware. This prototype provides an integrated view of the data stored in distributed heterogeneous relational databases through the Online Transaction Processing (OLTP) system of a data warehouse. Furthermore, this prototype provides the facility to distribute an SQL query through a data abstraction layer into multiple sub-queries aimed at the data marts containing the requested tables, and to combine the outcome of the individual sub-queries into a single consistent result.

The non-event data from the LHC will be generated at CERN and, like event data, will be distributed in multiple locations at sites around the world. Most of the data will be stored at the Tier-0 site at CERN, and at the seven Tier-1 sites. Smaller subsets of this data will be replicated to Tier-2 and Tier-3 sites when requested by scientists for analysis. Moreover, the database technologies used at the different tiers are also different. Oracle, for instance, is the most popular RDBMS system used at the Tier-0 and Tier-1 sites. On the other hand, MySQL and Microsoft SQL Server is the more common technology used at Tier-2 and Tier-3 sites. SQLite is the database favored by users who wish to do analysis while remaining disconnected over long periods of time (laptop users, for instance).

In order to manage these replicated sets of data, a system is required that can track the locations of the various databases, and provide efficient, transparent access to these datasets when queries are submitted to it by end users. In addition, with the growing linkages of Grid computing and Web service technologies, it is desirable to provide a Web service interface to the system so that client applications at every tier can access these services conveniently over the Web.

## 3. Related Work

The Grid is considered as a data-intensive infrastructure. Users expect the Grid to provide efficient and transparent access to enormous quantities of data scattered in globally distributed, heterogeneous databases. For this reason, integration of data retrieved from these databases becomes a major research issue.

Efforts have been made to solve this issue but performance bottlenecks still exist.

One of the projects targeting database integration is the Unity project [7] at the University of Iowa research labs. This project provides a JDBC driver, which uses XML specification files for integrating the databases. Then using the metadata information from the XML specification files, connections are established to the appropriate databases. The data is thus accessed without previous knowledge of the physical location of the data. Unity, however, does not do any load distribution, which causes some delays in query processing. As a result, if there is a lot of data to be fetched for a query, the memory becomes overloaded. In addition, it does not handle joins that span tables in multiple databases. In our work, we have used the Unity driver as the baseline for development. For our prototype, we have enhanced the driver with several features that are described in detail in Section 4.

Another project called the POOL Relational Abstraction Layer (POOL-RAL) [8], being pursued at CERN, provides a relational abstraction layer for relational databases and follows a vendor-neutral approach to database access. However, POOL provides access to tables within one database at a time, which puts a limit on the query and does not allow parallel execution of a query on multiple databases.

OGSA Distributed Query Processing (DQP) [9] is another project for distributed query processing on Grid-based databases. It distributes join operations on multiple nodes within a grid to take full advantage of the grid's distributed processing capabilities. However, OGSA-DQP is strongly dependent on the Globus Toolkit 3.2, which limits it to Globus only and makes it platform dependent.

IBM's Discovery Link [10] is another project aimed at carrying out integration of relational databases and other types of data sources for life sciences, genetics and bio-chemical data sources. However, due to the domain specific nature of this project, it cannot be used directly for HEP databases. ALDAP and SkyQuery are two other similar projects with the same basic objectives, but aimed at Astronomy databases.

# 4. System Architecture and Design

The distributed architecture of our system consists of the basic components described in the following discussion. Figure 1 shows an architectural diagram of the developed prototype.

The architecture consists of two main parts: The first part (the lower half of the diagram) retrieves data from various underlying databases, integrates it into a data warehouse, and replicates data from the data warehouse to the data marts, which are locally accessible by the client applications through the web-service interface. The second part (the upper half of the diagram) provides lightweight Clarens clients web service-based access to the data stored in the data marts.

## 4.1. Data Sources

The developed prototype supports Oracle and MySQL relational source databases. A normalized schema was developed for these source databases to store HBOOK [11] Ntuples data. The following example can help to understand the meaning of the Ntuples. Suppose that a dataset contains 10000 events and each event consists of many variables (say NVAR=200), then an Ntuple is like a table where these 200 variables are the columns and each event is a row. Furthermore, these databases were distributed over a Tier-1 center at CERN and a Tier-2 center at CALTECH.

## 4.2. Data Warehouse

The currently available approaches, as mentioned in section 3, provide different implementations to access data from distributed heterogeneous relational databases. Each approach provides a different implementation based on different driver requirements, connection requirements and database schema requirements of the databases. It means that for 'N' number of database technologies with 'S' number of database schemas, current approaches require 'NxS' number of implementations to be provided, in order to access data from globally distributed, heterogeneous databases. Furthermore, 'NxS' implementations also create a performance issue because each time access to data from these databases is requested, all the related meta-data information i.e. database schema, connection and vendor specific information, has to be parsed in order to return a reliable and consistent result of the query. In order to resolve this performance issue, we propose the use of a data warehouse.

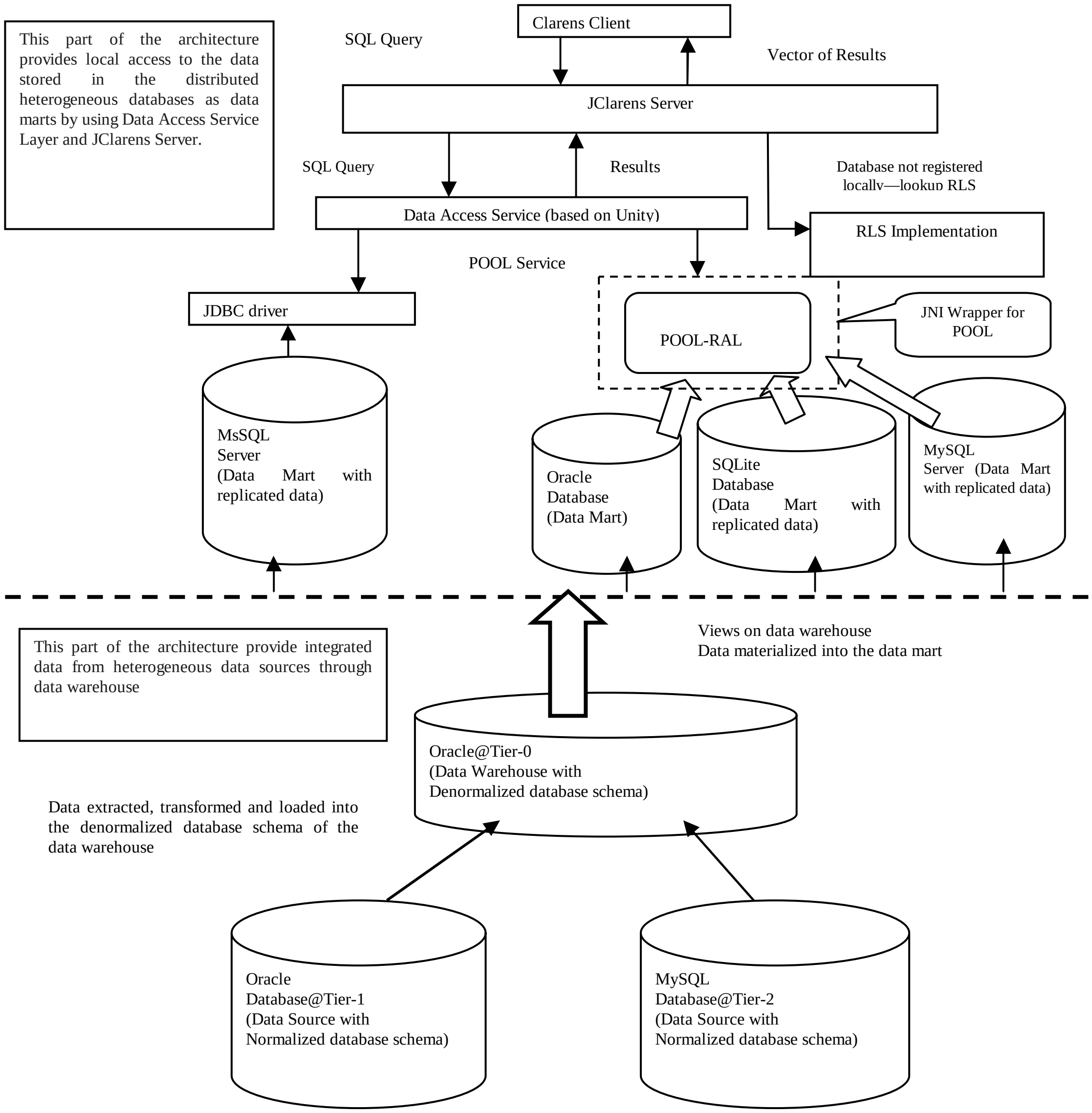

**Figure 1. Architectural Diagram.**

A data warehouse is a repository, which stores data that is integrated from heterogeneous data sources, for efficient querying, analysis and decision-making. Data warehouse technology is very successfully implemented in various commercial projects and is highly supported by vendors like ORACLE.

For the developed prototype, a denormalized star schema was developed in the ORACLE database. An Extraction, Transformation, Transportation and Loading (ETL) process was used to populate the data warehouse. In this ETL process, data was initially extracted from the distributed relational data sources, then integrated and transformed according to the denormalized database schema of the data warehouse. In this prototype, data streaming technology was used to perform the ETL process. Finally, this transformed data is loaded into the warehouse. In this prototype, we created views on the data stored in the warehouse to provide read-only access for scientific analysis.

## 4.3. Data Marts

A remote centralized data warehouse cannot be considered a good solution for an environment like the Grid, which is seen as a pool of distributed resources. In the context of databases, efficient accessibility of distributed databases can be achieved by making the required data available locally to the applications.

Thus, in order to utilize the features of the data warehouse successfully in a Grid environment without creating a centralized performance bottleneck, views are created on the integrated data of the data warehouse, and materialized on a new set of databases, which are made available locally to the applications. These databases are termed as data marts. Data marts are databases that store subsets of replicated data from the centralized data warehouse.

For the developed prototype, we create data marts, which are supported by MySQL, MS-SQL, ORACLE and SQLite. These databases are accessed using either through the POOL-RAL interface or using JDBC drivers, depending on whether or not they are supported by the POOL-RAL libraries.

## 4.4. XSpec Files

XSpec stands for "XML Specifications" files. These files are generated from the data sources using tools provided by the Unity project. Each database has its own XSpec file, which contains information about the schema of the database, including the tables, columns and relationships within the database. These logical names form a kind of data dictionary for the database, and this data dictionary is used for determining which database to access to fulfill a client's request. The client does not need to know the exact name of a database, tables in the database or names of the columns in the table. The client is provided this data dictionary of logical names, and he uses these logical names without any knowledge of the physical location of the data and their actual names. The query processing mechanism automatically maps logical names to physical names and divides the query to be executed among the individual databases.

These XSpec files are of two types:

**4.4.1. Lower Level XSpec.** The Lower Level XSpec refers to each individual database's XSpec file, which is generated from the original data source and contains the schema and all the other information mentioned above.

**4.4.2. Upper Level XSpec.** The Upper Level XSpec file is generated manually using the Lower Level XSpec files. This file just contains the URL for each database, the driver that each database is using and the name of the Lower Level XSpec for each database. There is only one Upper-Level XSpec file, whereas the number of lower-Level XSpec depends on the number of data sources.

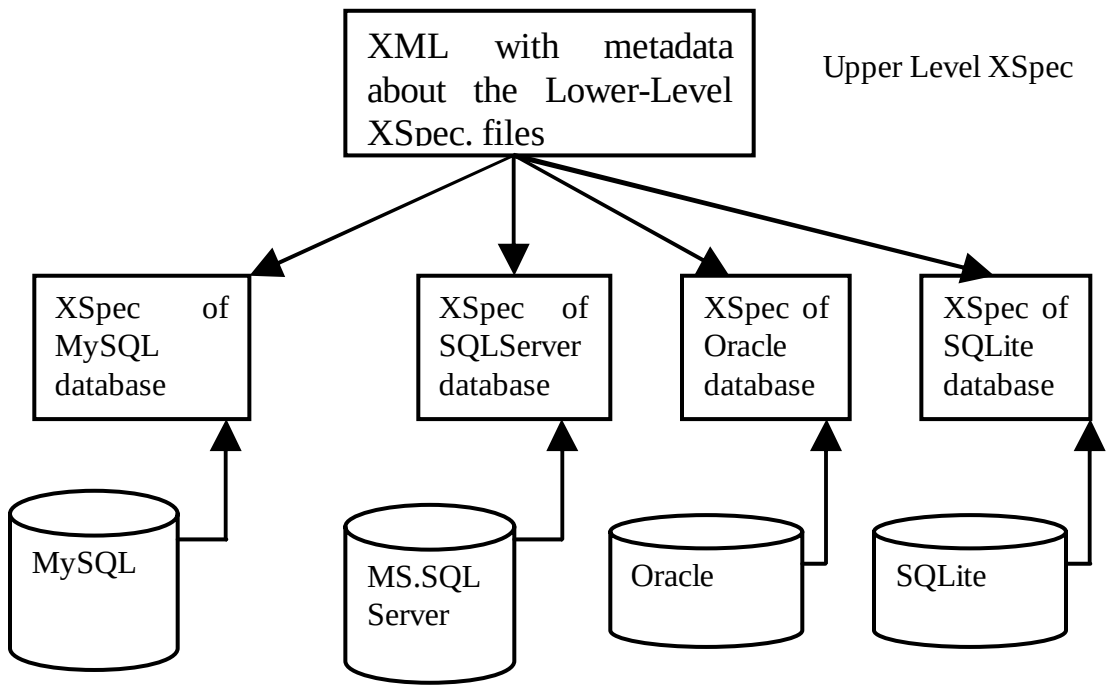

**Figure 2. Hierarchy of XSpec files**

## 4.5. Data Access Layer

This layer processes the queries for data sent by the clients containing joins of different tables from different databases (data marts), and divides them into sub-queries, which are then distributed on to the underlying databases.

The data access layer looks for the tables from which data is requested by the client. If the tables are locally registered with the JClarens server, the data access layer decides which of the two modules (POOL-RAL module or Unity driver) to forward the query to by finding out which databases are to be queried. If a database is supported by the POOL-RAL, the query is forwarded to the POOL RAL layer; otherwise, the query is forwarded to the JDBC driver. If the tables requested are not registered with the JClarens server, the Replica Location Service (RLS) is used to lookup the physical locations (hosting servers) of the tables. The RLS server provides the URL of the remote JClarens server with which the tables are registered. The queries are then forwarded to the remote servers, which perform the query processing, and send the retrieved data back to the original server, where the queries were submitted.

## 4.6. Unity Driver

As mentioned in Section 2 (Related Work) above, the Unity driver enables access to and integration of data from multiple databases. We have further enhanced this driver to be able to apply joins on rows extracted from multiple databases.

While accessing the underlying databases, the sub-queries meant for unsupported databases are accessed using the Unity driver, whereas the sub-queries concerned with POOL-supported databases are processed through the POOL RAL. The data retrieved through each of the sub-queries is finally merged into a single 2-D vector, and returned to the client.

### 4.7. POOL RAL Wrapper

The databases not supported by the POOL-RAL are handled by the JDBC driver. On the other hand, queries to databases supported by the POOL-RAL are forwarded through a wrapper layer to the POOL RAL libraries for execution.

The POOL RAL is implemented in C++ whereas JClarens and its services are implemented in Java. Therefore, to make the POOL libraries work with the JClarens based service, a JNI (Java Native Interface) wrapper was developed which exposes two methods:

1. One method initializes a service handler for a new database using a connection string, a username and a password and adds it to a list of previously initialized handles.
2. The other method takes as input a connection string, an array of select fields, an array of table names, and a 'where' clause string, and returns a 2D array containing the results of the query execution on the database represented by the connection string.

### 4.8. Replica Location Module

The Replica Location module is included in the project to distribute the load and reduce the query processing time, by enabling multiple instances of the database service to host smaller subsets from the entire collection of databases, and then collaborating with each other to provide access to one or more of those databases. In this way, load can be distributed over as many servers as required, instead of putting it entirely on just one server registering all the databases. This can also potentially enable us to achieve a hierarchical database hosting service in parallel with the tiered topology of the LHC Computing Model.

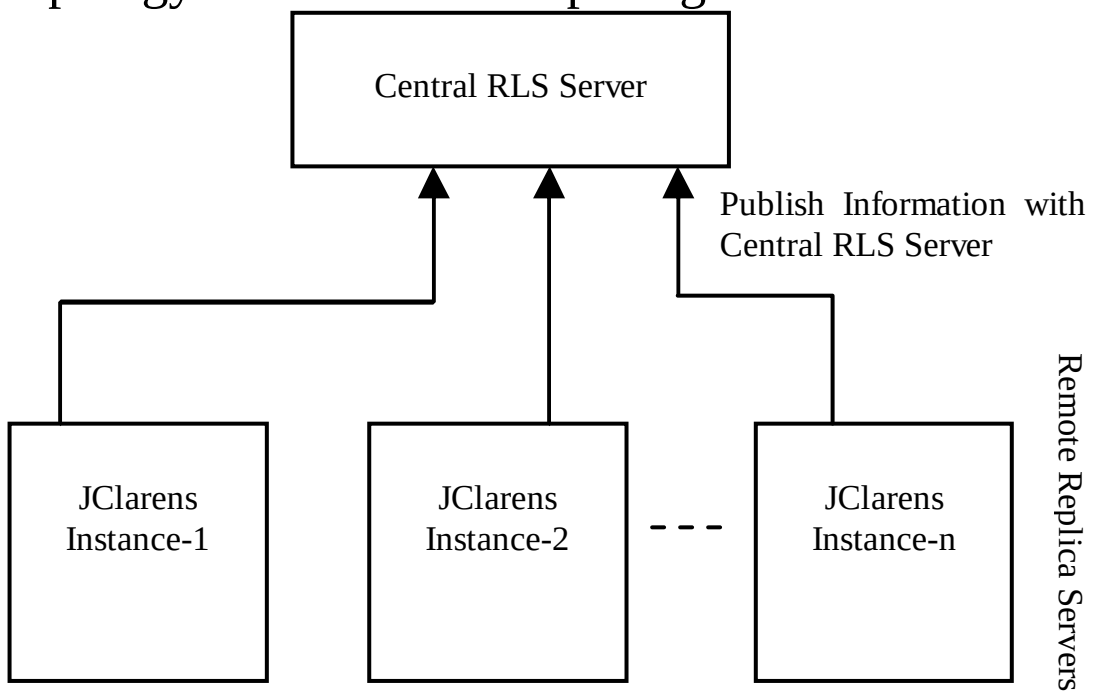

**Figure 3. Publishing table locations to the RLS**

This module uses a central RLS Server that contains the mapping of table names with replica servers' URLs. Each service instance publishes information about the databases and the tables it is hosting to the central RLS server. This central RLS server is contacted when the data access layer does not find a locally registered table.

### 4.9. Tracking Changes in Schema

The system is also able to track changes made to the schema of any database in the system. This feature enables the system to update itself according to changes in the schema of any of the databases.

The algorithm works as follows. After a fixed interval of time, a thread is run against the back-end databases to generate a new XSpec for each database. The size of the newly created XSpec is compared against the size of the older XSpec file. If the sizes are equal, the files are compared using their md5 sums. If there is any change in the size or md5 sum of the file, the older version of the XSpec is replaced by the new one. The JClarens server then uses the new XSpec file to update the schema it is using for that database.

### 4.10. Plug-in Databases

This feature enables databases to be added at runtime to the system. The server is provided the URL of the databases' XSpec file, the database driver name, and the database location. The server then downloads the file, parses it, and retrieves the metadata about the database. Using this metadata, the server establishes a connection with the database using the appropriate JDBC driver. When the connection is established, the server updates itself with the information about the tables contained in that database.

## 5. Performance Results

The developed prototype was tested in three stages:

Stage 1: Data is extracted from the source databases, transformed according to the denormalized schema requirements of the data warehouse, and then streamed into the data warehouse.

Stage 2: Data is extracted from the views, which were created on the data stored in the data warehouse, and materialized from the views (through data streaming) into the local databases i.e. data marts.

Stage 3: Response time of the distributed query, which runs through a JClarens based web interface, is measured against the locally available data marts.

### 5.1. Stage 1 and 2 results:

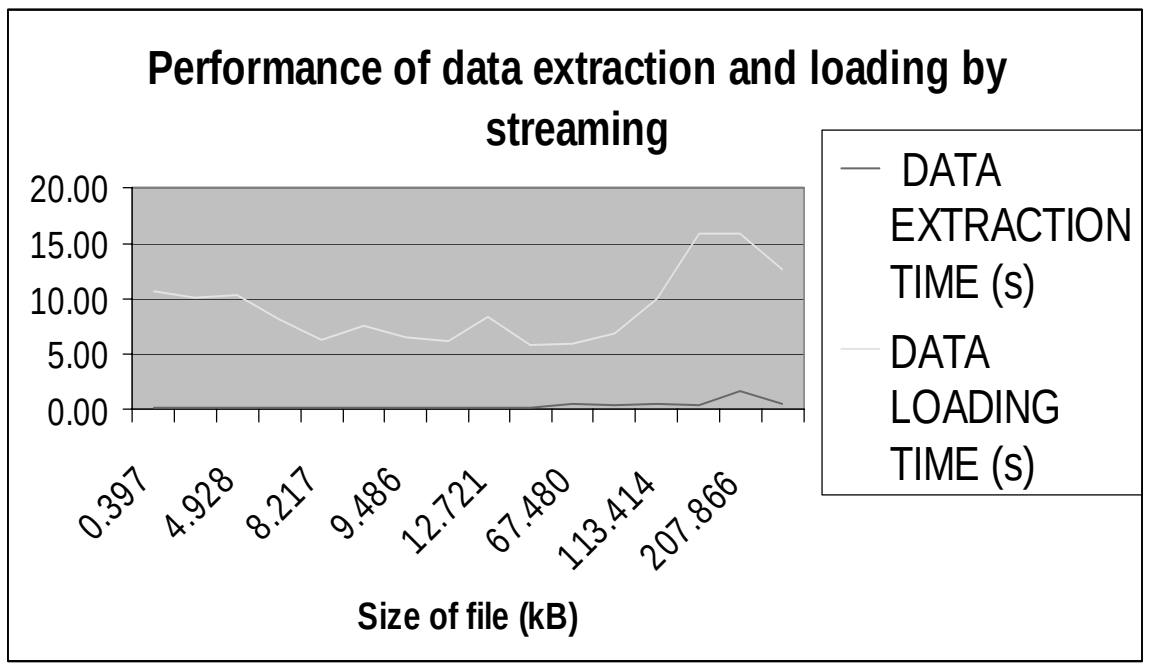

**Figure 4: Data extracted from source databases and loaded into the data warehouse.**

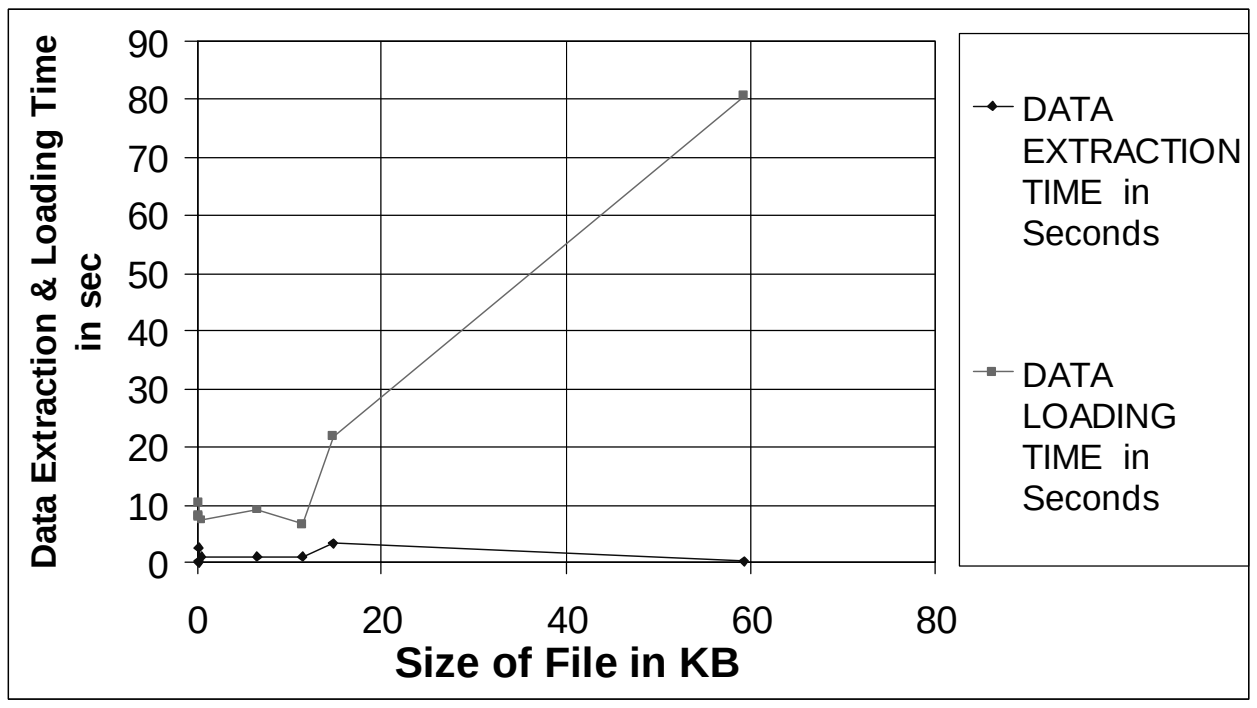

**Figure 5. Views extracted from the data warehouse and materialized into data marts.**

Stage 1 and 2 tests were carried out by streaming data of different sizes from data sources to the data warehouse and from data warehouse to the data marts respectively. The tests were carried out over a 100 Mbps Ethernet LAN. The respective data transfer time was plotted against the size of the transferred data. These plots, shown in figure 4 and 5, show the average data transfer time i.e. average of observations taken on different days at different time to measure the data transfer rate with different network traffic load. This time includes the time taken by a class to connect with the respective databases and, to open and close the stream for the respective SQL statements.

Each of the graphs shown in figure 4 and 5 are comprised of two plots, because in our prototype every time data was retrieved from a database it was first placed into a temporary file (data extraction) and then from this temporary file, data was stored into the other databases (data loading). In Figure 4, the lower line of the graph was plotted for the data extracted from the normalized data sources, transformed according to the denormalized schema of the data warehouse, and imported into a temporary file. The upper line shows the time taken to transfer data from the above generated temporary files to the data warehouse. Similarly, in Figure 5 the lower line of the graph shows the data retrieved from views, which were created on the data warehouse. The upper line shows the time taken to transfer the data from the generated temporary files and materialized into the data marts. Of course, the use of the temporary staging file during the process is a performance bottleneck, and we are working on a cleaner way of loading the warehouse directly from the normalized databases.

### 5.2. Stage 3 results:

We present here two aspects of the performance of the service. First, we measure the time the system takes to respond to a set of queries, each of which requires the involvement of a different number of Clarens servers and different number of databases. Secondly, we determine how the system throughput changes with different numbers of requested rows.

The tests were carried out on a 100 Mbps Ethernet LAN over two single-processor Intel Pentium IV (1.8 and 2.4 GHz) machines, with 512 MB and 1 GB of RAM respectively. The operating system on each machine was Redhat Linux 7.3. A Clarens server (with the data access service installed) was installed on each of the machines. The two servers were configured to host a total of 6 databases, with a total of nearly 80,000 rows and 1700 tables. The databases were equally shared between a Microsoft SQL Server on Windows 2000, and a MySQL database server.

**Table 1: Query Response Time**

| Number of Clarens servers accessed | Query Distributed (Yes/No) | Response Time | Number of tables accessed |
|---|---|---|---|
| 1 | No | 38 ms | 1 |
| 1 | Yes | 487.5 ms | 2 |
| 2 | Yes | 594 ms | 4 |

Table 1 (Query response time) shows the time in which the system responds to the client's queries. The column "Number of Clarens servers" shows the number of Clarens servers that had to be accessed in order to retrieve the requested rows of data. The "Query Distributed (Yes/No)" columns shows whether or not the query had to fetch data from multiple databases. The "Number of tables accessed" field represents the number of tables that were requested in that query. Although the response time for queries executing over multiple databases and servers are more than 10 times slower, it is inevitable because it involves determining which server to connect to using RLS, connecting and authenticating with several databases or servers, and integrating the results.

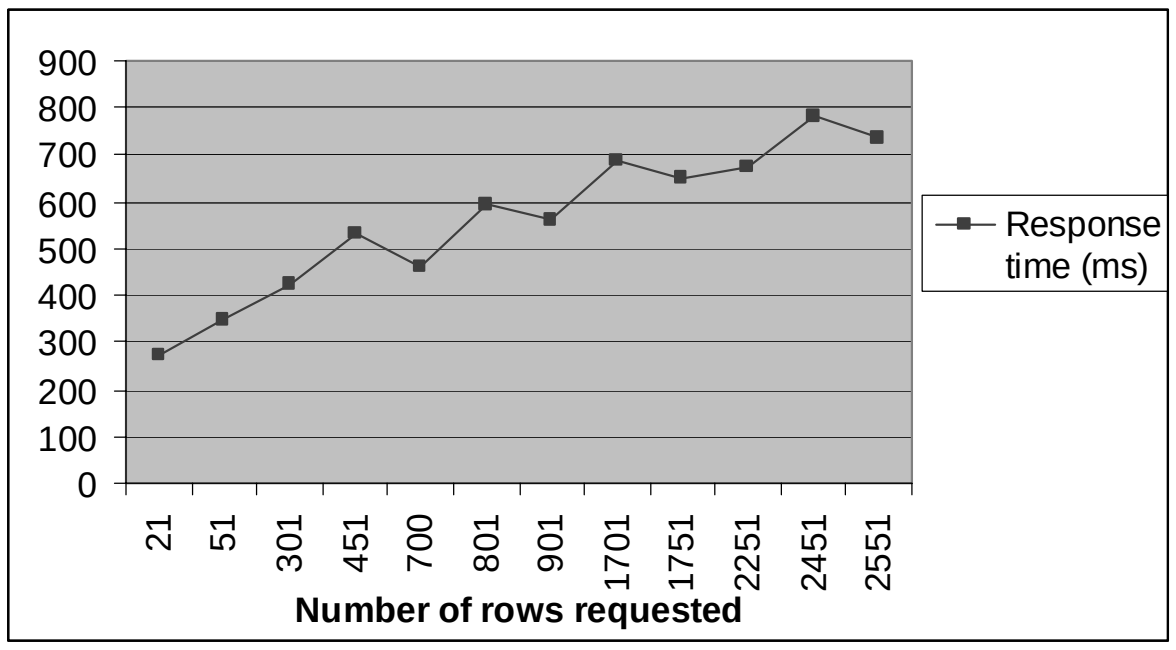

**Figure 6. Response time versus number of rows requested**

We also collected performance statistics to determine how the system scales with increasing number of rows requested by clients. For this purpose, we selected a number of queries to be run against the ntuple data in our databases, each of which was to return different number of rows. We determined the number of rows returned for each query, and measured the response time for each query to execute. The graph depicting the response time of the queries versus the number of rows returned is shown in Figure 6.

The graph shows that there is a linear increase in the response time of the system as a result of an increase in the number of requested rows. Increasing the number of rows from 21 to 2551 only increases the response time from about 300 to 700 ms. This shows that the system is scalable to support large queries. In addition, it is comparable to the performance reported by the OGSA-DAI project [12]. However, we are working on further improving the algorithms and implementation, to enable even better performance for very large queries.

## 6. Current Status and Future Work

A prototype has been developed, which is installable as an RPM on Redhat 7.3-based systems. The prototype possesses all of the features described above. However, some of the features such as joins spanning multiple databases have not been tested yet for all possible scenarios. Unit tests have been written for the system to check the integrity of the system. A plug-in for the Java Analysis Studio (JAS) [13] was also developed to submit queries for accessing the data and visualizing the results as histograms

Future directions will be to ensure the efficiency of the system and enhance the performance. In addition, we will be testing the system for query distribution on geographically distributed databases in order to measure its performance over wide area networks. We are also working on the design of a system that could decide the closest available database (in terms of network connectivity) from a set of replicated databases. Another interesting extension to the project could be the study of how tables from databases can be integrated with respect to their semantic similarity.

## 7. Conclusion

We have presented a system that enables heterogeneous databases distributed over the N-tiered architecture of the LHC experiment to present a single, simplified view to the user. With a single query, users can request and retrieve data from a number of databases simultaneously. This makes the (potentially) large number of databases at the backend transparent to the user while continuing to give them satisfactory performance.